\documentclass[12pt]{iopart}
\usepackage{graphicx}
\usepackage{bm}% bold math
\usepackage{amssymb}
\usepackage{setstack}
\begin{document}
%-----------------------------------------------
\title[Necessary and sufficient conditions for big bangs, bounces, crunches, rips...]
{Necessary and sufficient conditions for big bangs, bounces, crunches, rips, 
sudden singularities, and extremality events}%
%-----------------------------------------------

\author{%
C\'eline Catto\"en \footnote[1]{celine.cattoen@mcs.vuw.ac.nz},
and
Matt Visser \footnote[2]{matt.visser@mcs.vuw.ac.nz}}
\address{School of Mathematics, Statistics, and Computer Science, \\
Victoria University of Wellington, \\
P.O.Box 600, Wellington, New Zealand}

%-----------------------------------------------
\begin{abstract}
%-----------------------------------------------
The physically relevant  singularities occurring in FRW cosmologies had traditionally been thought to be limited to the ``big bang'', and possibly a ``big crunch''. However, over the last few years, the zoo of cosmological singularities considered in the literature has become  considerably more extensive, with ``big rips'' and ``sudden singularities'' added to the mix, as well as renewed interest in non-singular cosmological events such as ``bounces'' and ``turnarounds''. 
In this article we present a complete catalogue of such cosmological milestones, both at the kinematical and dynamical level. First, using generalized power series, purely kinematical definitions of these cosmological events are provided in terms of the behaviour of the scale factor $a(t)$. The notion of a ``scale-factor singularity'' is defined, and its relation to curvature singularities (polynomial and differential) is explored. Second, dynamical information is extracted by using the Friedmann equations (without assuming even the existence of any equation of state)  to place constraints on whether or not the classical energy conditions are satisfied at the cosmological milestones.  We use these considerations to derive necessary and sufficient conditions for the existence of cosmological milestones such as bangs, bounces, crunches, rips, sudden singularities, and extremality events. Since the classification is extremely general,  the corresponding results are to a high degree model-independent: In particular, we provide a complete characterization of the class of bangs, cruncjes, and sudden singularities for which the dominant energy condition is satisfied.

\bigskip

\centerline{gr-qc/0508045; 30 August 2005; \LaTeX-ed \today}

%-----------------------------------------------
\end{abstract}
%-----------------------------------------------


%-----------------------------------------------
\maketitle
%-----------------------------------------------

\def\d{{\mathrm{d}}}
\def\union{\cup}
\def\implies{\Rightarrow}
%------------------------------
\def\bang{{\ast}}
\def\crunch{{\circledast}}
%--------------------------
\def\bounce{{\bullet}}
\def\inflexion{{\circledcirc}}
\def\turnaround{{\circ}}
%---------------------------
\def\rip{{\divideontimes}}
%---------------------------
\def\sudden{{\circleddash}}
%---------------------------
\def\generic{{\odot}}
%-------------------------------------------------------------------------
%--------------------------------------------------------------------------

%-----------------------------------------------
\section{Introduction}
%-----------------------------------------------
The cosmological principle, which is ultimately a distillation of our knowledge of observational cosmology, leads one to consider cosmological spacetimes of the idealized FRW form~\cite{MTW, Wald, Carroll, Hartle}:
\begin{equation}
\d s^{2}=-\d t^{2}+a(t)^{2}\left\{\frac{\d r^{2}}{1-kr^{2}}+r^{2}\;[\d\theta^{2}+\sin^2\theta\; \d\phi^{2}]\right\}.
\end{equation}
The central question in cosmology is now the prediction, or rather retro-diction, of the history of the scale factor $a(t)$~\cite{jerk1,jerk2}.  The key dynamical equations (assuming applicability of the Einstein equations of general relativity to cosmology in the large, an assumption which at least some cosmologists are beginning to question) are the Friedmann equations (which redundantly imply the conservation of stress-energy). Indeed, in units where $8\pi G_N=1$ and $c=1$, we have
\begin{equation}
 \rho(t)=3\left(\frac{\dot a^{2}}{a^{2}}+\frac{k}{a^{2}}\right),
\label{Friedmann1}
\end{equation}
\begin{equation}
p(t)=-2\,{\ddot a\over a}-{\dot a^2\over a^2} - {k\over a^2},
\label{Friedmann2}
\end{equation}
\begin{equation}
\rho(t)+3\,p(t)=-6\,{\ddot a\over a},
\label{Friedmann3}
\end{equation}
and the related conservation equation
\begin{equation}
 \dot\rho(t)\;a^{3}+3\,[\rho(t)+p(t)]\,a^{2}\,\dot a=0,
\label{conservation}
\end{equation}
where any two of equations (\ref{Friedmann1}, \ref{Friedmann2}, \ref{Friedmann3}) imply (\ref{conservation}).~\footnote{If in contrast you try to use equations (\ref{conservation}) and one of  (\ref{Friedmann1}, \ref{Friedmann2}, \ref{Friedmann3})  to derive the other two equations you may need to make some specific choice of integration constants.   For instance,  (\ref{conservation}) and (\ref{Friedmann3}) do not contain $k$, and in this situation $k$ arises as an integration constant, which can without loss of generality be set to $k= -1/0/+1$ by suitably rescaling $a(t)$. Secondly, and more significantly,   combining  (\ref{conservation}) and (\ref{Friedmann2}) reproduces (\ref{Friedmann1}) only up to an ambiguity, $\rho \sim K/a^3$, corresponding to an arbitrary quantity of ``dust'' which has to be eliminated ``by hand''.} Until quite recently the only FRW cosmological singularities receiving significant attention were the big bang (and its time-reversed partner, the big crunch) where $a(t)\to 0$. Recently, the class of interesting cosmological singularities has become much broader~\cite{rip, rip2, rip-details, quiescent, sudden1, sudden2, sudden3, sudden-details, Lake}. In this article we shall pin down necessary and sufficient conditions for the various cosmological singularities and related events, (which we shall generically refer to as cosmological milestones), both at the level of kinematics, and at the level of dynamics. 
\begin{itemize}
\item The kinematical analysis boils down to developing precise definitions of these cosmological milestones (``major events'') in terms of the scale factor $a(t)$.
\item The dynamical analysis uses the Friedmann equations to analyze the classical energy conditions in the vicinity of these milestones in a model-independent manner (without even assuming the existence of any equation of state).
\end{itemize}
The overarching aim of the article is to provide a (hopefully complete) catalogue of the interesting cosmological milestones that might be encountered in unusual cosmological models, and determine (with an absolute minimum of technical assumptions) the necessary and sufficient conditions for their occurrence.

%-----------------------------------------------
\section{Kinematics --- definitions}
%-----------------------------------------------

\paragraph{Generic cosmological milestone:} Suppose we have some unspecified generic cosmological milestone,  that  is defined in terms of the behaviour of the scale factor $a(t)$, and which occurs at some finite time $t_\generic$. We will assume that in the vicinity of the milestone the scale factor has a (possibly one-sided) generalized power series expansion of the form
\begin{equation}
a(t) = c_0 |t-t_\generic|^{\eta_0} + c_1  |t-t_\generic|^{\eta_1} + c_2  |t-t_\generic|^{\eta_2} 
+ c_3 |t-t_\generic|^{\eta_3} +\dots
\end{equation}
where the indicial exponents $\eta_i$ are generically real (but are often non-integer) and without loss of generality are ordered in such a way that they satisfy
\begin{equation}
\eta_0<\eta_1<\eta_2<\eta_3\dots
\end{equation}
Finally we can also without loss of generality set
\begin{equation}
c_0 > 0.
\end{equation}
There are no \emph{a priori} constraints on the signs of the other $c_i$, though by definition $c_i\neq0$.

Physically, this generalized power series expansion is sufficient to encompass all the models we are aware of in the literature, and the indicial exponents $\eta_i$ will be used to classify the type of cosmological milestone we are dealing with. Indeed the lowest few of the indicial exponents are sufficient to determine the relationship between these cosmological milestones and the curvature singularities more traditionally encountered in the general relativity literature. 
Mathematically, power series of this type generalize the notions of Taylor series, Laurent series, and Frobenius series, and are even more general than the generalized Frobenius series adopted in~\cite{power-laws}.  The current definition is motivated in part by the fact that generalized Frobenius series (compositions of power-law prefactors with Taylor series in some variable raised to a possibly different exponent)  commonly occur when expanding solutions of differential equations around their  singular points. 

For many of our calculations the first term in the expansion is dominant, but even for the most subtle of the explicit calculations below  it will be sufficient to keep only the first three terms of the expansion:
\begin{equation}
\fl
a(t) = c_0 |t-t_\generic|^{\eta_0} + c_1  |t-t_\generic|^{\eta_1} + c_2  |t-t_\generic|^{\eta_2} \dots;
\qquad
\eta_0 < \eta_1<\eta_2; 
\qquad c_0 > 0.
\end{equation}

\paragraph{Big bangs and big crunches:}

The most basic of the cosmological milestones are big bangs and big crunches, for which the scale factor $a(t)\to0$ at some finite time as we move to the past or future. Let the time of the big bang (if one occurs) be denoted by $t_\bang$ and the time of the big crunch (if one occurs) be denoted by $t_\crunch$. We shall say that the bang or crunch behaves with indicial exponents ($0<\eta_0<\eta_1\dots$) if the scale factor possesses a generalized power series in the vicinity of the singularity. That is, if
\begin{eqnarray}
a(t) =  c_0 (t-t_\bang)^{\eta_0} + c_1 (t-t_\bang)^{\eta_1} +\dots
\end{eqnarray}
or
\begin{eqnarray}
a(t) =  c_0 (t_\crunch-t)^{\eta_0} \; + c_1 (t_\crunch-t)^{\eta_1} + \dots
\end{eqnarray}
respectively.  Note that the series have been carefully constructed to make $a(t_\bang)=0$ and $a(t_\crunch)=0$.

\paragraph{Big rips:} A ``big rip'' is said to occur if $a(t)\to\infty$ at finite time~\cite{rip,rip2}. We can distinguish a ``future rip'' from a ``past rip'', where the literature to date has solely considered future rips (as a past rip would be a most unusual and unexpected beginning to the history of the universe). Let the time of the rip, if it occurs, be denoted $t_{\rip}$, then we define the indicial exponents of the rip (either future or past) to be
($\eta_0<\eta_1\dots$) if the scale factor possesses a generalized power series in the vicinity of the rip:
\begin{eqnarray}
a(t) = c_0  |t_{\rip}-t|^{\eta_0} + c_1  |t_{\rip}-t|^{\eta_1} + \dots,
\end{eqnarray}
with $\eta_0<0$ and $c_0\neq0$.
Note the similarity to bangs and crunches, with the only difference being in the \emph{sign} of the exponent $\eta_0$. Note that the series has been carefully constructed to make $a(t_\rip)=\infty$.  

\paragraph{Sudden singularities:} Past or future sudden singularities are a recent addition to the collection of cosmological milestones, and are defined by some time derivative of the scale factor diverging at finite time, while the scale factor itself remains finite~\cite{sudden1, sudden2, sudden3}. Again, almost all attention has been confined to future sudden singularities as (a past sudden singularity would be a most unusual and disturbing beginning to the history of the universe). Let the time of the sudden singularity, if one occurs, be $t_{\sudden}$ (past or future). A suitable definition of the exponent of a sudden singularity is to take $\eta_0=0$ and $\eta_1>0$ to give
\begin{eqnarray}
a(t) =  c_0  + c_1 |t-t_\sudden|^{\eta_1} + \dots
\end{eqnarray}
with  $c_0>0$ and $\eta_1$ non-integer. Thus $a(t_\sudden)=c_0$ is finite and a sufficient number of differentiations yields~\footnote{In particular the toy model considered by Barrow~\cite{sudden1, sudden2, sudden3} can be written 
\begin{equation}
a(t) = c_0 \left[ (t_\sudden-t)^\eta -1 \right] + \tilde c_0 (t-t_\bang)^{\tilde\eta}
\end{equation}
and falls into this classification when expanded around the time of the sudden singularity, $t_\sudden$,  while it falls into the classification of big bang singularities considered above when expanded around the time of the big bang, $t_\bang$.} 
\begin{equation}
\fl
a^{(n)}(t\to t_\sudden) \sim c_0 \; \eta_1 (\eta_1-1)(\eta_1-2)\dots (\eta_1-n+1) \; |t-t_\sudden|^{\eta_1-n}\to\infty.
\end{equation}

\paragraph{Extremality events:} Other common cosmological milestones are extremality events, which are not singularities in any sense. For these events $a(t)$ exhibits a local extremum at finite time, so in particular $\dot a \to 0$. In the vicinity of extremality events we can model the scale factor using ordinary Taylor series so that in terms of our generalized series we have $\eta_0=0$ and $\eta_i\in Z^+$. In particular
\begin{itemize}
\item A ``bounce'' is any local minimum of $a(t)$, the time of such an event being denoted by $t_\bounce$, so that  $a^{(1)}(t_\bounce)=0$ is zero~\cite{bounce0, bounce, Tolman, bounce-details}. The ``order'' of the bounce is the first nonzero integer $n$ for which the $2n$'th time derivative is strictly positive~\cite{bounce,Tolman}:
\begin{equation}
a^{(2n)}(t_\bounce) > 0,
\end{equation}
so that 
\begin{equation}
a(t) = a(t_\bounce) + {1\over(2n)!} a^{(2n)}(t_\bounce)\; [t-t_\bounce]^{2n} + \dots.
\end{equation}

\item A ``turnaround'' is any local maximum of $a(t)$, the time of such an event being denoted by $t_\turnaround$, 
so that  $a^{(1)}(t_\turnaround)=0$ is zero~\cite{recollapse}. The ``order'' of the turnaround is the first nonzero integer $n$ for which the $2n$'th time derivative is strictly negative:
\begin{equation}
a^{(2n)}(t_\turnaround) < 0,
\end{equation}
so that
\begin{equation}
a(t) = a(t_\turnaround) + {1\over(2n)!} a^{(2n)}(t_\turnaround)\; [t-t_\turnaround]^{2n} + \dots.
\end{equation}

\item An ``inflexion event'' is an extremality event that is neither a local maximum or a local minimum, the time of such an event being denoted by $t_\inflexion$. (Inflexion events can be thought of as an extreme case of ``loitering''~\cite{loitering}, in the limit where the Hubble parameter momentarily vanishes at the inflexion event.)
The order of the inflexion event is the first nonzero $n$ for which
\begin{equation}
a^{(2n+1)}(t_\inflexion) \neq 0,
\end{equation}
so that
\begin{equation}
a(t) = a(t_\inflexion) + {1\over(2n+1)!} a^{(2n+1)}(t_\inflexion)\; [t-t_\inflexion]^{2n+1} + \dots.
\end{equation}

\item The ``emergent universe'' of~\cite{emergent} can be thought of as an extremality event that has been pushed back into the infinite past.

\end{itemize}
These definitions have been chosen to match with and simplify the definitions in articles~\cite{bounce} and ~\cite{Tolman}.
Note that for bounces these definitions imply that there will be some open interval such that
\begin{equation}
\forall t \in (t_\bounce-\Delta,t_\bounce)\union (t_\bounce,t_\bounce+\Delta); \qquad \ddot a(t) > 0,
\end{equation}
while for turnarounds there will be some open interval such that
\begin{equation}
\forall t \in (t_\turnaround-\Delta,t_\turnaround)\union  (t_\turnaround,t_\turnaround+\Delta); \qquad \ddot a(t) < 0.
\end{equation}
Note that unless the bounce or turnaround is of order one we cannot guarantee that at the extremality event itself $\ddot a(t_\bounce)>0$ or $\ddot a(t_\turnaround)<0$.
For inflexion events we can only assert the weaker condition of the existence of  some open interval such that
\begin{equation}
\forall t \in(t_\inflexion-\Delta,t_\inflexion)\union (t_\inflexion,t_\inflexion+\Delta); \qquad
\dot a(t) \hbox{ has fixed sign}.
\end{equation}

\paragraph{Summary:} With these definitions in place we can put all of the singular cosmological milestones (bangs, crunches, rips, sudden singularities), and the non-singular cosmological milestones (extremality events)  into a single framework based on generalized power series --- and we see purely on kinematic grounds that major items of interest will be the first two indicial exponents ($\eta_0$ and $\eta_1$).  Indeed $\eta_0>0$ for bangs and crunches, $\eta_0<0$ for rips, and $\eta_0=0$ for sudden singularities (and also, with additional conditions, for extremality events).
For the nonsingular milestones (bounces, inflexions, turnarounds) the parameterization can be  somewhat simpler, needing only one positive integer --- the order of the extremality event, to specify the qualitative behaviour of the Taylor series $a(t)$. 
In the next two sections we will use these parameters to explore the generic properties of the cosmological milestones, from the presence (or absence) of curvature singularities, to the violation (or otherwise) of the energy conditions.

%-----------------------------------------------
\section{Kinematics --- spacetime curvature}
%-----------------------------------------------

To analyze the spacetime curvature, let us first consider the Hubble parameter $H=\dot a/a$.
From the 
definition of a generic cosmological milestone
\begin{equation}
a(t) = c_0 |t-t_\generic|^{\eta_0} + c_1  |t-t_\generic|^{\eta_1} +\dots
\end{equation}
we have (assuming $t>t_\generic$ for simplicity, otherwise one need merely formally reverse the flow of time)\footnote{In fact for explicit calculations in the vicinity of any cosmological milestone it is always possible to choose the direction of time to force  $t>t_\generic$ and so dispense with the need to take the absolute value $|t-t_\generic|$, at least for one-sided calculations. This is not a physical restriction on the  cosmological milestone, just a mathematical convenience which we shall adopt henceforth without further explicit discussion.}
\begin{equation}
\dot a(t) = c_0 \eta_0 (t-t_\generic)^{\eta_0-1} + c_1  \eta_1 (t-t_\generic)^{\eta_1-1}  +\dots
\end{equation}
Keeping only the most dominant term, we have for $\eta_0\neq0$
\begin{equation}
H ={\dot a\over a} 
\sim 
{c_0 \eta_0 (t-t_\generic)^{\eta_0-1}\over c_0 (t-t_\generic)^{\eta_0}} 
=  
{\eta_0\over t-t_\generic};  \qquad (\eta_0\neq0).
\end{equation}
That is, for bangs, crunches, and rips the Hubble parameter exhibits a generic $1/(t-t_\generic)$ blow up. For $\eta_0=0$ (corresponding to either a sudden singularity or an extremality event) we need to go to the next highest term in the numerator (a term which depends on $\eta_1$, which is guaranteed to be greater than zero by our definitions) to obtain
\begin{equation}
H \sim  {c_1 \eta_1 (t-t_\generic)^{\eta_1-1}\over c_0 } =  
\eta_1 \;{c_1\over c_0} \; (t-t_\generic)^{\eta_1-1};
\qquad (\eta_0=0; \eta_1>0).
\end{equation}
In particular, this guarantees that power law behaviour is completely generic near the cosmological milestone, and while  the value of the exponent is typically $-1$, there is an exceptional class of milestones (the sudden singularities and extremality events) for which the exponent will differ. Note that it is \emph{not} automatic that the Hubble parameter diverge at the cosmological milestone. Indeed 
\begin{equation}
\lim_{t\to t_\generic} H = \left\{ \begin{array}{ll}
+\infty & \eta_0>0;\\
\mathrm{sign}(c_1)\infty  & \eta_0=0; \qquad \eta_1\in(0,1);\\
c_1/c_0 & \eta_0=0; \qquad \eta_1=1;\\
0 & \eta_0=0; \qquad \eta_1 > 1;\\
-\infty &\eta_0<0.\\
\end{array}
\right.
\end{equation}
In particular the Hubble parameter has a finite limit iff $\eta_0=0$, $\eta_1\geq1$, corresponding to a particular subset of the sudden singularities.
We have presented this in some detail because the style of argument is generic. Consider for instance the cosmological acceleration $\ddot a$:
\begin{equation}
\fl
\ddot a(t) = c_0 \eta_0 (\eta_0-1) |t-t_\generic|^{\eta_0-2} 
+ c_1  \eta_1 (\eta_1-1) |t-t_\generic|^{\eta_1-2}  
+ c_2  \eta_2  (\eta_2-1) |t-t_\generic|^{\eta_2-2}  
+\dots
\end{equation}
Then provided $\eta_0\neq 0$ and $\eta_0\neq 1$
\begin{equation}
{\ddot a\over a} \sim {\eta_0(\eta_0-1)\over (t-t_\generic)^2}.
\end{equation}
If $\eta_0=0$, then provided $\eta_1\neq 1$
\begin{equation}
{\ddot a\over a} \sim {\eta_1(\eta_1-1)c_1\over c_0} \;  (t-t_\generic)^{\eta_1-2}.
\end{equation}
If both $\eta_0=0$ and $\eta_1= 1$
\begin{equation}
{\ddot a\over a} \sim {\eta_2(\eta_2-1)c_2\over c_0} \;  (t-t_\generic)^{\eta_2-2}.
\end{equation}
Finally, if $\eta_0=1$, then (since $\eta_1>1$)
\begin{equation}
{\ddot a\over a} \sim {\eta_1(\eta_1-1)c_1\over c_0} \;  (t-t_\generic)^{\eta_1-3}.
\end{equation}
Again the generic message is that the behaviour of $\ddot a/a$ near the milestone will be \emph{some} power law, though the precise exponent of that power law will depend on the interplay between the various indicial exponents $\eta_i$. Note that there is at least one situation in which we have to calculate up  to the third exponent $\eta_2$.
We can now consider the so-called deceleration parameter
\begin{equation}
q = - {\ddot a\over a} H^{-2} = - {a\; \ddot a\over \dot a^2}
\end{equation}
Following the same sort of analysis, for $\eta_0\neq0$ and $\eta_0\neq1$ we have the ``generic" result:
\begin{equation}
q \sim  {1-\eta_0\over \eta_0};  \qquad (\eta_0\neq 0,1).
\end{equation}
For the ``exceptional'' cases we easily see:\\
--- If $\eta_0=0$, then provided $\eta_1\neq 1$
\begin{equation}
q \sim -{(\eta_1-1)c_0\over \eta_1 c_1} \;  (t-t_\generic)^{-\eta_1}.
\end{equation}
--- If both $\eta_0=0$ and $\eta_1= 1$
\begin{equation}
q \sim -{c_2 c_0\over c_1^2} \eta_2(\eta_2-1) \;  (t-t_\generic)^{\eta_2-2}.
\end{equation}
--- Finally, if  $\eta_0=1$, then (since $\eta_1>1$)
\begin{equation}
q \sim - {\eta_1(\eta_1-1)c_1\over c_0} \;  (t-t_\generic)^{\eta_1-1}.
\end{equation}
Again we see the ubiquity of power law behaviour, with a ``generic'' case and a limited number of special cases. Note that the deceleration parameter can be either finite or infinite as the cosmological milestone  is approached:
\begin{equation}
\fl
\lim_{t\to t_\generic} q = \left\{ \begin{array}{ll}
(1-\eta_0)/\eta_0 & \qquad\eta_0\neq0;\\
\mathrm{sign}(c_1[1-\eta_1])\infty & \qquad\eta_0=0; \qquad \eta_1\neq 1;\\
0 & \qquad\eta_0=0; \qquad \eta_1= 1; \qquad\eta_2 > 2;\\
-2c_2c_0/c_1^2& \qquad\eta_0=0; \qquad \eta_1= 1; \qquad\eta_2=2; \\
-\mathrm{sign}(c_2)\infty &\qquad \eta_0=0; \qquad \eta_1= 1; \qquad\eta_2\in(1,2). \\
\end{array}
\right.
\end{equation}
Note that the deceleration parameter has an infinite limit only for a certain subset of the sudden singularities $\eta_0=0$. For bangs and crunches the limit is always finite.\footnote{Of course this is largely because the definition of the deceleration parameter was carefully chosen to eliminate the leading $t$ behaviour whenever possible.}

\def\Rtt{{ R_{\hat t\hat t} }}
\def\Gtt{{ G_{\hat t\hat t} }}

\paragraph{Polynomial curvature singularities:} 
To now formally decide whether the cosmological milestones we have defined (bangs, crunches, rips, sudden singularities, extremality events) are curvature singularities in the sense commonly encountered in general relativity we will have to look  at the Riemann tensor. Because of the symmetries of FRW geometry, the Weyl tensor is automatically zero and so it suffices to consider the Ricci tensor (and implicitly the Ricci scalar). But because of spherical symmetry, and the perhaps less obvious translational symmetry, the only two non-zero orthonormal  independent components of the Ricci tensor are
\begin{equation}
R_{\hat t\hat t} \qquad\hbox{and} \qquad R_{\hat r\hat r}=R_{\hat\theta\hat\theta}=R_{\hat\phi\hat\phi}.
\end{equation}
Thus to test for all possible polynomial curvature singularities it suffices to test for singularities in, for instance, $R_{\hat t\hat t}$ and $R_{\hat r\hat r}$. Alternatively one could consider $\Rtt$ and the Ricci scalar $R$, or even $\Rtt$ and $\Gtt$. Indeed a brief computation reveals that
\begin{equation}
\Rtt = - 3 \;{\ddot a\over a};
\end{equation}
\begin{equation}
\Gtt = 3 \left( {\dot a^2\over a^2} + {k\over a^2}\right).
\end{equation}
In fact these two particular linear combinations of the orthonormal components of the Ricci tensor satisfy the useful properties that:
\begin{itemize}
\item they are linearly independent;
\item $\Rtt$ is independent of the curvature of space (no $k$ dependence);
\item $\Gtt$ is independent of $\ddot a$ (minimizing the number of derivatives involved);
\item testing these two objects for finiteness is sufficient to completely characterize \emph{all} polynomial curvature singularities in a FRW geometry. 
\end{itemize}
This is observation already enough to tell us that almost all of the cosmological milestones defined above are in fact polynomial curvature singularities. Indeed the number of situations in which $\Rtt$ remains finite is rather limited. The generic result is
\begin{equation}
\lim_{t\to t_\generic} \Rtt = \mathrm{sign}(\eta_0[1-\eta_0])\, \infty
\qquad \hbox{for} \qquad \eta_0\neq 0; \quad \eta_0\neq 1,
\end{equation}
which focuses attention on $\eta_0=0$ and $\eta_0=1$.
From our previous results we see that 
\begin{equation}
\fl
\lim_{t\to t_\generic} \Rtt = \left\{ \begin{array}{ll}
-\infty & \qquad\eta_0>1; \\
0 & \qquad\eta_0=1; \qquad \eta_1>3;\\
-18c_1/c_0& \qquad\eta_0= 1; \qquad\eta_1=3; \\
-\mathrm{sign}(c_1)\infty &\qquad \eta_0= 1; \qquad\eta_1\in(0,3); \\
+\infty & \qquad\eta_0\in(0,1); \\
0 & \qquad\eta_0=0;  \qquad \eta_1 > 2;\\
-6c_1/c_0 & \qquad\eta_0=0;  \qquad \eta_1 = 2;\\
\mathrm{sign}(c_1)\infty & \qquad\eta_0=0; \qquad 
\eta_1\in(1,2);\\
0 & \qquad\eta_0=0; \qquad \eta_1= 1; \qquad\eta_2 > 2;\\
-6c_2/c_0& \qquad\eta_0=0; \qquad \eta_1= 1; \qquad\eta_2=2; \\
-\mathrm{sign}(c_2)\infty &\qquad \eta_0=0; \qquad \eta_1= 1; \qquad\eta_2\in(1,2); \\
-\mathrm{sign}(c_1)\infty & \qquad\eta_0=0; \qquad  \eta_1 \in(0,1);\\
-\infty & \qquad \eta_0<0.
\end{array}
\right.
\end{equation}
Therefore $\Rtt$ is finite provided
\begin{itemize}
\item $\eta_0=0$, $\eta_1\geq 2$;
\item $\eta_0=0$, $\eta_1=1$, $\eta_2\geq2$;
\item $\eta_0=1$ and $\eta_1\geq 3$.
\end{itemize}
To analyze $\Gtt$ recall that the condition that the Hubble parameter $H$ remain finite was $\eta_0=0$, $\eta_1 \geq 1$. Thus if $k\geq0$ then 
$\Gtt$ will remain finite iff $H$ remains finite.  In contrast, if $k<0$ then $\Gtt$ is a difference of squares and one term can be balanced against the other. Indeed near the milestone
\begin{equation}
\Gtt \sim 3 \left[ {\eta_0^2\over(t-t_\generic)^2} + {k\over c_0^2(t-t_\generic)^{2\eta_0}} \right]
\qquad \eta_0\neq0,
\end{equation}
while
\begin{equation}
\Gtt \sim 3 \left[ {\eta_1^2c_1^2(t-t_\generic)^{2(\eta_1-1)} + k\over c_0^2} \right]
\qquad \eta_0=0.
\end{equation}
After a brief calculation 
\begin{equation}
\fl
\lim_{t\to t_\generic} \Gtt = \left\{ \begin{array}{ll}
\mathrm{sign}(k)\infty & \eta_0 > 1; \qquad k\neq 0;\\
+\infty & \eta_0 > 1; \qquad k= 0;\\
\mathrm{sign}(c_0^2+k)\infty & \eta_0 = 1; \qquad c_0^2+k\neq0;\\
0 &\eta_0=1; \qquad c_0^2+k=0; \qquad \eta_1>3;\\
18 c_1 &\eta_0=1; \qquad c_0^2+k=0; \qquad \eta_1=3; \\
\mathrm{sign}(c_1)\infty &\eta_0=1; \qquad c_0^2+k=0; \qquad \eta_1\in(1,3);\\
+\infty & \eta_0 \in(0,1);\\
3k/c_0^2 & \eta_0=0;  \qquad \eta_1 > 1;\\
3(c_1^2+k)/c_0^2 & \eta_0=0; \qquad \eta_1=1.\\
+\infty & \eta_0=0; \qquad \eta_1\in(0,1).\\
\end{array}
\right.
\end{equation}
Thus the necessary and sufficient conditions for $\Gtt$ to remain finite at the cosmological milestone are:
\begin{itemize}
\item $\eta_0=0$, $\eta_1\geq 1$;
\item $\eta_0=1$, $c_0^2+k=0$, and  $\eta_1\geq 3$.  \\
(Note that $c_0^2+k=0$ implies $k=-1$, and $c_0=1$.)
\end{itemize}
Combining with the previous result,
the necessary and sufficient conditions for both $\Gtt$ and $\Rtt$ to remain finite, so that a cosmological milestone is \emph{not} a polynomial curvature singularity, are:
\begin{itemize}
\item $\eta_0=0$, $\eta_1\geq 2$;
\item $\eta_0=0$, $\eta_1=1$, $\eta_2\geq2$;
\item $\eta_0=1$, $k=-1$, $c_0=1$, and  $\eta_1\geq 3$.
\end{itemize}
This can only happen for a particular sub-class of sudden singularities ($\eta_0=0$), and for a rather exceptional type of big bang/crunch, one that asymptotes to a Milne universe (see discussion below). However, once one generalizes the concept of polynomial curvature singularity to that of a derivative curvature singularity, most of these exceptional cases will be excluded. 

\paragraph{Derivative curvature singularities:}
A derivative curvature singularity is defined by some polynomial constructed from finite-order derivatives of the curvature tensor blowing up. Because of the symmetries of the FRW universe the only interesting derivatives will be time derivatives, and so the only objects we need consider are
\begin{equation}
{\d^n \Rtt\over \d^n t} \qquad \hbox{and} \qquad {\d^n \Gtt\over \d^n t}.
\end{equation}
But
\begin{equation}
{\d^n \Rtt\over \d^n t} = - 3 {a^{(n+2)}\over a} + \hbox{ (lower-order derivatives)};
\end{equation}
while
\begin{equation}
 {\d^n \Gtt\over \d^n t} = 3 {\dot a \; a^{(n+1)}\over a^2}  + \hbox{ (lower-order derivatives)}.
\end{equation}
Thus to avoid a $n$th-order derivative curvature singularity we must at the very least keep $a^{(n+2)}/a$ finite, and furthermore all related lower-order derivatives of the form $a^{(j)}/a$,  with $j\leq n+2$, must also be finite. 
To prevent any arbitrary-order derivative singularity from occurring, we must force all  $a^{(j)}/a$ to remain finite.  This condition holds in addition to the constraint coming from polynomial curvature singularities derived above.

Since the constraint from polynomial curvature singularities had already forced $\eta_0=0$ or $\eta_0=1$, then this implies that we must force all  $a^{(j)}$ to remain finite, and so must force all
the indicial exponents $\eta_i$ to be non-negative integers, thus making $a(t)$ a Taylor series.  This condition holds in addition to the constraint coming from polynomial curvature singularities derived above, with the consequence that almost all cosmological milestones are indeed curvature singularities. Indeed, inserting the condition that $a(t)$ be representable by a Taylor series into $\Gtt$ and $\Rtt$ shows that  the \emph{only} two situations in which a cosmological milestone is \emph{not} a derivative curvature singularity is if:
\begin{itemize}

\item $\eta_0=0$,\; $\eta_{i}\in Z^+$; corresponding to an extremality event (bounce, turnaround, or inflexion event) rather than a bang, crunch, rip, or sudden singularity;

\item $\eta_0=1$, $k=-1$, $c_0=1$, $\eta_{i}\in Z^+$, and  $\eta_1\geq 3$; corresponding to a FRW geometry that smoothly asymptotes near the cosmological milestone to the Riemann-flat  Milne universe. (The special case $a(t)=t$, $k=-1$ is called the Milne universe and is actually a disguised portion of Minkowski space. It corresponds to the interior of the future light cone based at some randomly specified point in Minkowski space, with a spatial foliation defined by the proper time hyperboloids based on that event. Since this spacetime is flat, it corresponds to a universe which on the largest scales is empty. This is not a popular cosmological model.)
\end{itemize}
In other words, almost all cosmological milestones are physical singularities, apart from a very limited sub-class corresponding to either extremality events (bounces, turnarounds, inflexion events) or an asymptotically empty universe.
It is important to realise that up to this stage our considerations have been purely \emph{kinematic} --- up to this point we have not used or needed the Friedmann equations or the Einstein equations.

%***************************************

%-----------------------------------------------
\section{Dynamics --- energy conditions}
%-----------------------------------------------

To  now start to include dynamics we relate the geometry to the density and pressure using the Friedmann equations and then ask what happens to the energy conditions at the cosmological milestones. Even though we may have good reason to suspect that the energy conditions are not truly fundamental~\cite{twilight} --- they make a very good first pass at the problem of quantifying just how ``strange'' physics gets at the cosmological milestone. The standard energy conditions are the \emph{null}, \emph{weak}, \emph{strong}, and \emph{dominant} energy conditions which for a FRW spacetime specialise to~\cite{bounce, Tolman, twilight}:
\begin{itemize}

\item{}[NEC] $\rho+p \geq 0$. \\
In view of the Friedmann equations this reduces to
\begin{equation}
\ddot a \leq {\dot a^2+k \over a}; \qquad \hbox{that is} \qquad k \geq a \;\ddot a - \dot a^2.
\end{equation}

\item{}[WEC] This specialises to the NEC plus $\rho\geq0$.\\
This reduces to the NEC plus the condition
\begin{equation}
\dot a^2 + k \geq 0; \qquad \hbox{that is} \qquad k \geq - \dot a^2.
\end{equation}

\item{}[SEC] This specilalises to the NEC plus $\rho+3p\geq0$.\\
This reduces to the NEC plus the deceleration condition
\begin{equation}
\ddot a \leq 0.
\end{equation}

\item{}[DEC] $\rho\pm p \geq 0$.\\
This reduces to the NEC plus the condition
\begin{equation}
\ddot a \geq -{2(\dot a^2+k) \over a}; \qquad \hbox{that is}
\qquad k \geq -{(a\; \ddot a +2\dot a^2)\over2}.
\end{equation}
\end{itemize}
Of these, it is the NEC that is the most interesting (because this is the weakest of the standard energy conditions it leads to the strongest theorems), and we shall analyze this condition in most detail.

\paragraph{NEC:}
Now near any generic cosmological milestone
\begin{equation}
a \;\ddot a - \dot a^2 \sim - \eta_0 \;c_0^2 \;t^{2(\eta_0-1)}; \qquad (\eta_0\neq0),
\end{equation}
while in the degenerate cases
\begin{equation}
a \;\ddot a - \dot a^2 \sim 
c_0 c_1 \; \eta_1(\eta_1-1) \; (t-t_\generic)^{(\eta_1-2)};  \qquad (\eta_0=0; \eta_1\neq1),
\end{equation}
and
\begin{equation}
a \;\ddot a - \dot a^2 \sim 
c_0 c_2 \; \eta_2(\eta_2-1) \; (t-t_\generic)^{(\eta_2-2)}  - c_1^2; \qquad (\eta_0=0; \eta_1=1).
\end{equation}
Therefore
\begin{equation}
\fl
\lim_{t\to t_\generic} (a\;\ddot a-\dot a^2)  = 
\left\{ \begin{array}{ll}
0  & \qquad \eta_0>1;\\
-c_0^2 & \qquad\eta_0=1;\\
-\infty &\qquad \eta_0 \in(0,1);\\
 0 & \qquad \eta_0=0; \qquad \eta_1>2;\\
 2 c_0 c_1 & \qquad \eta_0=0; \qquad \eta_1=2;\\
+  \mathrm{sign}(c_1) \infty &\qquad \eta_0=0; \qquad \eta_1\in(1,2)\\
 -c_1^2 & \qquad \eta_0=0; \qquad \eta_1=1; \qquad \eta_2>2;\\
 2c_0c_2 -c_1^2 & \qquad \eta_0=0; \qquad \eta_1=1; \qquad \eta_2=2;\\
+  \mathrm{sign}(c_2) \infty   & \qquad \eta_0=0; \qquad \eta_1=1; \qquad \eta_2\in(1,2);\\
- \mathrm{sign}(c_1) \infty &\qquad \eta_0=0; \qquad \eta_1\in(0,1)\\
 +\infty &\qquad \eta_0 < 0.\\
\end{array}
\right.
\end{equation}
Hence the NEC is \emph{definitely satisfied} (meaning the inequality is \emph{strict}) at a generic cosmological milestone iff:
\begin{itemize}
\item $\eta_0>1$, $k=+1$;
\item $\eta_0=1$: for $k=+1$ and $k=0$, and also for $k=-1$ with the proviso that $c_0> 1$;
\item $\eta_0\in(0,1)$: for any value of $k$; 
\item $\eta_0=0$ subject to the additional constraints:
\begin{itemize}
\item $\eta_1>2$ and $k=+1$;
\item $\eta_1=2$ and $k>2c_0c_1$;
\item $\eta_1\in(1,2)$ and $c_1<0$;
\item $\eta_1=1$, $\eta_2>2$: for $k=+1$, $k=0$, and for $k=-1$ with the proviso $|c_1|>1$;
\item $\eta_1=1$, $\eta_2=2$: for $k>2c_0c_2-c_1^2$;
\item $\eta_1=1$, $\eta_2\in(1,2)$, and $c_2<0$;
\item $\eta_1\in(0,1)$, and $c_1>0$.
\end{itemize}
\end{itemize}
Hence the NEC is \emph{marginally satisfied} (meaning the non-strict inequality is actually an equality) at a generic cosmological milestone iff:
\begin{itemize}
\item $\eta_0>1$, $k=0$;
\item $\eta_0=1$, $k=-1$ with the proviso that $c_0= 1$;
\item $\eta_0=0$ subject to the additional constraints:
\begin{itemize}
\item $\eta_1>2$ and $k=0$;
\item $\eta_1=2$ and $k=2c_0c_1$ (which requires $k\neq0$);
\item $\eta_1=1$, $\eta_2>2$: for $k=-1$ and $c_1=\pm1$;
\item $\eta_1=1$, $\eta_2=2$, and $k=2c_0c_2-c_1^2$.
\end{itemize}
\end{itemize}
In all other situations the NEC is definitely violated. To make this explicit, the NEC is definitely violated (the inequality is strictly violated) at a generic milestone iff:
\begin{itemize}
\item $\eta_0>1$, $k=-1$;
\item $\eta_0=1$: for $k=-1$ with the proviso that $c_0< 1$;
\item $\eta_0=0$ subject to the additional constraints:
\begin{itemize}
\item $\eta_1>2$ and $k=-1$;
\item $\eta_1=2$ and $k<2c_0c_1$;
\item $\eta_1\in(1,2)$ and $c_1<0$;
\item $\eta_1=1$, $\eta_2>2$:  for $k=-1$ with the proviso $|c_1|<1$;
\item $\eta_1=1$, $\eta_2=2$: for $k<2c_0c_2-c_1^2$;
\item $\eta_1=1$, $\eta_2\in(1,2)$, and $c_2>0$;
\item $\eta_1\in(0,1)$, and $c_1<0$.
\end{itemize}
\item $\eta_0 < 0$.
\end{itemize}
This pattern is in agreement with the standard folklore, and is a systematic expression of results that are otherwise scattered throughout the literature.  For instance, it is immediate that ``big rips'' \emph{always} violate the NEC (and hence all the other energy conditions), and that for big bangs and crunches the range $\eta_0\in(0,1)$ is preferred, since it is only in this range that the NEC holds \emph{independent} of the spatial curvature.  Note that for sufficiently violent big bangs ($\eta_0>1$) and hyperbolic spatial curvature the NEC is violated --- this indicates that ``phantom matter'' need not always lead to a ``big rip"~\cite{rip,rip2}; it might in contrast lead to a particularly violent bang or crunch.
It is for $\eta_0=0$ (which corresponds to either sudden singularities or extremality events) that the analysis becomes somewhat tedious. Certainly, many (though not all) of the sudden singularities violate the NEC.

\paragraph{WEC:}
If in addition we consider the WEC, we need to look at
\begin{equation}
 - \dot a^2 \sim - \eta_0^2 c_0^2 (t-t_\generic)^{2(\eta_0-1)}; \qquad (\eta_0\neq0),
\end{equation}
while in the degenerate case
\begin{equation}
 - \dot a^2 \sim - \eta_1^2 c_1^2 (t-t_\generic)^{2(\eta_1-1)}; \qquad (\eta_0=0).
\end{equation}
Therefore
\begin{equation}
\fl
\lim_{t\to t_\generic} (-\dot a^2)  = 
\left\{ \begin{array}{ll}
0  & \qquad \eta_0>1;\\
-c_0^2 & \qquad\eta_0=1;\\
-\infty &\qquad \eta_0 \in(0,1);\\
 0 & \qquad \eta_0=0; \qquad \eta_1>1;\\
 -c_1^2 & \qquad \eta_0=0; \qquad \eta_1=1;\\
- \infty &\qquad \eta_0=0; \qquad \eta_1\in(0,1)\\
 -\infty &\qquad \eta_0 < 0.\\
\end{array}
\right.
\end{equation}
Since we want the condition $k>-\dot a^2$ to hold in addition to the NEC, it follows that the WEC
is equivalent to the NEC for bangs crunches and rips, and that the only changes arise when dealing with $\eta_0=0$ (sudden singularities or extremality events).  Certainly, many (though not all) of the sudden singularities violate the WEC.

\paragraph{SEC:}
When analyzing the SEC, it is sufficient to note that
near any generic cosmological milestone:
\begin{equation}
\ddot a  \sim \eta_0(\eta_0-1) \;c_0 \;t^{(\eta_0-2)}; \qquad (\eta_0\neq0, \;\;\; \eta_0\neq 1),
\end{equation}
while in the degenerate cases
\begin{equation}
\ddot a  \sim 
\eta_1(\eta_1-1) \; c_1 \;  (t-t_\generic)^{(\eta_1-2)};  \quad 
\left\{ \begin{array}{ll}
\eta_0=0; \qquad \eta_1\neq1; \\
\eta_0=1;
 \end{array}\right.
\end{equation}
\begin{equation}
\ddot a  \sim 
\eta_2(\eta_2-1) \; c_2 \;  (t-t_\generic)^{(\eta_2-2)};  \qquad (\eta_0=0; \;\;\; \eta_1=1; ),
\end{equation}
Therefore
\begin{equation}
\lim_{t\to t_\generic^+} [\mathrm{sign}(\ddot a)] = 
\left\{ \begin{array}{ll}
+1  & \qquad \eta_0>1\;\\
\mathrm{sign}(c_1) & \qquad \eta_0=1;\\
-1 & \qquad \eta_0\in(0,1):\\
\mathrm{sign}(\eta_1[\eta_1-1] c_1 ) & \qquad\eta_0=0;\qquad  \eta_1\neq 1;\\
\mathrm{sign}(c_2)  &\qquad \eta_0 =0; \qquad \eta_1=1;\\
 +1 &\qquad \eta_0 < 0.\\
\end{array}
\right.
\end{equation}
Since we want the condition $\ddot a \leq 0$ to hold in addition to the NEC, it follows that the SEC is definitely violated for all ``rips'', and is also definitely violated for all ``violent'' 
bangs and crunches with $\eta_0>1$. The SEC is definitely satisfied for bangs and crunches in the range $\eta_0\in(0,1)$. For turnarounds or bounces  $\eta_1=2n$, $n\in Z^+$, and the \emph{sign} of $c_1$ governs possible SEC violations: Turnarounds satisfy the SEC while bounces violate the SEC.  (See also~\cite{bounce,Tolman}.) For inflexion events,  where  $\eta_1=2n+1$, $n\in Z^+$, the SEC is violated either just before or just after the inflexion event.
(See also~\cite{bounce,Tolman}.) 

\paragraph{DEC:} 
Finally, when analyzing the DEC, it is sufficient to note that
near any generic cosmological milestone:
\begin{equation}
\fl
-{(a \ddot a  + 2 \dot a^2)\over 2} \sim 
-{\eta_0(3\eta_0-1) \;c_0^2\over 2}
 \;(t-t_\generic)^{2(\eta_0-1)}; \qquad (\eta_0\neq0, \;\;\; \eta_0\neq 1/3),
\end{equation}
while in the various degenerate cases
\begin{equation}
\fl
-{(a \ddot a  + 2 \dot a^2)\over 2} \sim 
-{\eta_1(\eta_1-1) \;c_0 \; c_1\over 2}
 \;(t-t_\generic)^{(\eta_1-2)};
 \qquad 
(\eta_0=0; \;\;\; \eta_1\neq1),
\end{equation}
\begin{equation}
\fl
-{(a \ddot a  + 2 \dot a^2)\over 2} \sim 
-{\eta_2(\eta_2-1) \;c_0 \; c_2\over 2}
 \;(t-t_\generic)^{(\eta_2-2)}-c_1^2;
\qquad (\eta_0=0; \;\;\; \eta_1=1),
\end{equation}
\begin{equation}
\fl
-{(a \ddot a  + 2 \dot a^2)\over 2} \sim 
-{(3\eta_1+2)(3\eta_1-1) \;c_0 \; c_1\over 18}
 \;(t-t_\generic)^{(\eta_1-5/3)};
\qquad (\eta_0=1/3).
\end{equation}
Therefore
\begin{equation}
\fl
\lim_{t\to t_\generic} 
\left[-{(a \ddot a  + 2 \dot a^2)\over 2} \right]
= 
\left\{ \begin{array}{ll}
0  & \qquad \eta_0>1\;\\
- c_0^2 & \qquad\eta_0=1;\\
-\infty &\qquad \eta_0 \in(1/3,1);\\
%
%%%
0 & \qquad \eta_0=1/3; \qquad \eta_1>5/3;\\
-{14\over9} c_0 c_1 & \qquad \eta_0=1/3; \qquad \eta_1=5/3;\\
-\mathrm{sign}(c_1)\infty & \qquad \eta_0=1/3; \qquad \eta_1<5/3;\\
+\infty &\qquad \eta_0 \in(0,1/3);\\
%%%%
0 & \qquad \eta_0=0; \qquad \eta_1>2;\\
%%%%%
-c_0c_1 & \qquad \eta_0=0; \qquad \eta_1=2;\\
%%%%%
-\mathrm{sign}(c_1)\infty   & \qquad \eta_0=0; \qquad \eta_1\in(1,2)\\
%%%%%
-c_1^2 & \qquad \eta_0=0; \qquad \eta_1=1; \qquad \eta_2 >2;\\
%%%%%
-c_0c_2 - c_1^2& \qquad \eta_0=0; \qquad \eta_1=1; \qquad \eta_2=2;\\
%%%%%
-\mathrm{sign}(c_2)\infty & \qquad \eta_0=0; \qquad \eta_1=1; \qquad \eta_2\in(1,2);\\
%%%%%
+\mathrm{sign}(c_1)\infty & \qquad \eta_0=0; \qquad \eta_1\in(0,1);\\
%%%%%
-\infty &\qquad \eta_0 <0.\\
\end{array}
\right.
\end{equation}
Remember that to satisfy the DEC one needs to satisfy the NEC in addition to the constraint coming from the above. Let us define
\begin{equation}
K = \mathrm{max} \left\{   
\lim_{t\to t_\generic} 
\left[ a \ddot a - \dot a^2 \right],
\lim_{t\to t_\generic} 
\left[-{(a \ddot a  + 2 \dot a^2)\over 2} \right]  
\right\}.
\end{equation}
Then satisfying the DEC at the cosmological milestone is equivalent to the constraint
\begin{equation}
k \geq K.
\end{equation}
--- For bangs and crunches we calculate:
\begin{equation}
K = 
\left\{ \begin{array}{ll}
0  & \qquad \eta_0>1;\\
- c_0^2 & \qquad\eta_0=1;\\
-\infty &\qquad \eta_0 \in(1/3,1);\\
%
%%%
0 & \qquad \eta_0=1/3; \qquad \eta_1>5/3;\\
-{14\over9} c_0 c_1 & \qquad \eta_0=1/3; \qquad \eta_1=5/3;\\
-\mathrm{sign}(c_1)\infty & \qquad \eta_0=1/3; \qquad \eta_1<5/3;\\
+\infty &\qquad \eta_0 \in(0,1/3).\\
\end{array}
\right.
\end{equation}
In particular, if one wishes to use ``normal'' matter (that is, matter satisfying all of the energy conditions) to drive a bang or a crunch independent of the sign of space curvature $k$, then one is forced in a model independent manner into the range $\eta_0\in(1/3,1)$. 
\\
--- For rips we have:
\begin{equation}
K = +\infty \qquad \eta_0 <0.
\end{equation}
Note that rips, because they violate the NEC, always violate the DEC, so the particular constraint derived above is not the controlling feature. 
\\
--- Finally, for sudden singularities and extremality events:
\begin{equation}
\fl
K = 
\left\{ \begin{array}{ll}
%%%%%
0 & \qquad \eta_0=0; \qquad \eta_1>2;\\
%%%%%
c_0\;\mathrm{max}\{2c_1,-c_1\} & \qquad \eta_0=0; \qquad \eta_1=2;\\
%%%%%
+\infty   & \qquad \eta_0=0; \qquad \eta_1\in(1,2)\\
%%%%%
-c_1^2 & \qquad \eta_0=0; \qquad \eta_1=1; \qquad \eta_2 >2;\\
%%%%%
c_0\;\mathrm{max}\{2c_1,-c_1\} - c_1^2& \qquad \eta_0=0; \qquad \eta_1=1; \qquad \eta_2=2;\\
%%%%%
+\infty & \qquad \eta_0=0; \qquad \eta_1=1; \qquad \eta_2\in(1,2);\\
%%%%%
+\infty & \qquad \eta_0=0; \qquad \eta_1\in(0,1).\\
%%%%%
%
%
\end{array}
\right.
\end{equation}
As we have now seen (several times), the $\eta_0=0$ case is ``special'' and requires extra care and delicacy in the analysis --- this is the underlying reason why ``sudden singularities'' are so ``fragile'', and so dependent on the specific details of the particular model. Indeed there are several classes of cosmological milestone for which the DEC is \emph{satisfied}. For instance, a complete catalogue of the bangs and crunches for which the DEC is satisfied is:
\begin{itemize}
\item $\eta_0>1$, $k=0, +1$.
\item $\eta_0=1$,  $k=0, +1$, and $k=-1$ if $c_0\geq 1$.
\item $\eta_0\in(1/3,1)$, for all $k$.
\item $\eta_0=1/3$, $\eta_1>5/3$, $k=0, +1$.
\item  $\eta_0=1/3$, $\eta_1=5/3$, $k\geq-{14\over9}c_0c_1$.
\item  $\eta_0=1/3$, $\eta_1\in(1/3,5/3)$, $c_1>0$, for all $k$.
\end{itemize}
In particular, if the DEC is to hold independent of the sign of space curvature $k$ all the way down to the singularity, then in the vicinity of the singularity the dominant term in the scale factor $a(t)$ is bounded from both above and below by
\begin{equation}
c_0 (t-t_\bang)^{1/3} \leq a_\mathrm{dominant}(t) < c_0 (t-t_\bang).
\end{equation}
Similarly, a complete catalogue of the sudden singularities for which the DEC is satisfied is:
\begin{itemize}
\item $\eta_0=0$, $\eta_1>2$, $k=0, +1$.
\item $\eta_0=0$, $\eta_1=2$, with $k\geq c_0\;\mathrm{max}\{2c_1,-c_1\}$.\\
(In particular this requires $k=+1$ though that is not sufficient.)
\item  $\eta_0=0$, $\eta_1=1$, $\eta_2>2$, $k=0, +1$, and $k=-1$ if $|c_1|\geq1$.
\item  $\eta_0=0$, $\eta_1=1$, $\eta_2=2$, $k\geq c_0\;\mathrm{max}\{2c_1,-c_1\} - c_1^2$.
\end{itemize}
This \emph{appears} at first glance, to contradict the analysis of Lake~\cite{Lake}, who claimed that \emph{all} sudden singularities violate the DEC.  The resolution of this apparent contradiction lies in the question ``\emph{just how sudden is the sudden singularity?}'' Lake takes as his definition $\ddot a(t\to t_\generic) = -\infty$, corresponding in our analysis to $\eta_0=0$ with either $\eta_1\in(0,1)\cup(1,2)$ or $\eta_1=1$ with $\eta_2<2$. In this situation our results certainly agree with those of Lake --- if $\ddot a(t\to t_\generic) = -\infty$ then the DEC is certainly violated. However, our  ``counterexamples'' where the DEC is satisfied all satisfy $\ddot a(t\to t_\generic) = \hbox{finite}$, and for these counterexamples it is only some higher derivative  $a^{(n)}(t\to t_\generic)$ for $n\geq3$ that diverges. Sudden singularities of this type are sufficiently ``gentle'' that at least for some of them even the DEC can be satisfied all the way to the singularity --- a conclusion completely in agreement with Barrow and Tsagas~\cite{sudden3}, though now we have a complete characterization of those situations for which the DEC can be satisfied.\footnote{See also the discussion by Sahni and Shtanov~\cite{quiescent}  in a braneworld context.}

%-----------------------------------------------
\section{Discussion}
%-----------------------------------------------

In this article we have explored three issues: First we have developed a complete catalogue of the various ``cosmological milestones'' in terms of a generalized power series expansion of the FRW scale factor in the vicinity of the milestone. This power series expansion is sufficiently general to accommodate all commonly occurring models considered in the literature. 

Second, with the notion of a generalized power series in hand, it is possible at a purely kinematic level to address the question of when a ``cosmological milestone'' corresponds to a curvature singularity, and what type of singularity is implied. In this way is is possible to classify all ``cosmlogical milestones'' as to whether they are singular or not, with the class non-singular milestones being strictly limited (to extremality events or asymptotically Milne bangs/crunches). 

Third, this definition of cosmological milestones in terms of generalized power series enables us to perform a complete model-independent check on the validity or otherwise of the classical energy conditions. In particular we provide a complete catalogue of those bangs/crunches, sudden singularities and extremality events for which the DEC is \emph{satisfied}, and a complete catalogue of those bangs/crunches, sudden singularities and extremality events for which the NEC is \emph{satisfied}.
Depending on one's attitude towards the energy conditions~\cite{twilight}, one could use this catalogue as a guide towards deciding on potentially interesting scenarios to investigate. 

Finally, we remind the reader of what is perhaps the most general lesson to be learned: If in the vicinity of any cosmological milestone, the input scale factor $a(t)$ is a generalized power series, then all physical observables ($H$, $q$, the Riemann tensor, \emph{etc}.) will likewise be generalized power series, with related indicial exponents that can be calculated from the indicial exponents of the scale factor. Whether or not the particular physical observable then diverges at the cosmological milestone is ``simply'' a matter of calculating its dominant indicial exponent in terms of those occuring in the scale factor.

%-----------------------------------------------
\section*{Acknowledgements}
%-----------------------------------------------

This research was supported by the Marsden Fund administered by the Royal Society of New Zealand.

%-----------------------------------------------
\section*{References}
%-----------------------------------------------

%-----------------------------------------------

%-----------------------------------------------

%-----------------------------------------------
\end{document}